\documentclass[prb,aps,twocolumn,amsmath,amsfonts]{revtex4}

\usepackage{graphicx}

\def\Tr{\mbox{Tr}\,}

\begin{document}

\title{On the statistics of quantum transfer of non-interacting
fermions in multi-terminal junctions}

\author{Dania Kambly}
\affiliation{Institute of Theoretical Physics,
Ecole Polytechnique F\'ed\'erale de Lausanne (EPFL), 
CH-1015 Lausanne, Switzerland}

\author{D.~A.~Ivanov}
\affiliation{Institute of Theoretical Physics,
Ecole Polytechnique F\'ed\'erale de Lausanne (EPFL), 
CH-1015 Lausanne, Switzerland}

\date{October 7, 2009}

\begin{abstract}
Similarly to the recently obtained result for two-terminal
systems, we show that there are constraints on the full counting
statistics for non-interacting fermions in multi-terminal
contacts. In contrast to the two-terminal result, however,
there is no factorization property in the multi-terminal case.
\end{abstract}

\maketitle

{\it Introduction. }
The problem of full counting statistics (FCS) of electronic charge
transfer has been addressed since 
long time,\cite{1993-LevitovLesovik} and the particular model of
non-interacting fermions has been studied in detail in various 
setups. The FCS for
transfer of non-interacting fermions is given by the Levitov--Lesovik
determinant formula\cite{1993-LevitovLesovik,1993-IvanovLevitov,%
1996-LevitovLeeLesovik,1997-IvanovLeeLevitov,2002-Klich} valid at arbitrary
temperature and for an arbitrary time evolution of the scatterer.
Recently, some properties of this result have been elucidated. First,
in the particular case of charge transfer driven by a time-dependent
bias voltage at zero temperature, the resulting FCS enjoys certain
symmetries.\cite{2007-VanevicNazarovBelzig,2008-VanevicNazarovBelzig,2008-AbanovIvanov}
Second, in the more general case of an arbitrary time-dependent
scatterer and at arbitrary temperature, it has been shown that
the FCS is factorizable into independent single-particle 
events.\cite{2008-AbanovIvanov,2009-AbanovIvanov}

In the present work, we generalize the  result of 
Refs.~\onlinecite{2008-AbanovIvanov,2009-AbanovIvanov}
to a multi-terminal setup. As in those works, we address the
problem of determining which multi-channel
charge transfers are possible and which are not in an arbitrary
quantum pump, in the model of non-interacting fermions.
In the two-terminal case, the constraint derived in
Refs.~\onlinecite{2008-AbanovIvanov,2009-AbanovIvanov}
is exact. In the multiterminal case, however, 
the problem is more complicated, and we have only partially solved it: 
we have formulated a {\it necessary} constraint  (a ``convexity condition'')
on the charge-transfer statistics, without a proof (or a counterexample) 
that this constraint is sufficient. Also, there is no obvious physical 
interpretation of this constraint: we show that, unlike in the two-terminal 
case, our constraint cannot be interpreted as a factorization property of the
charge-transfer statistics. 

This work is partly based on the results reported
in Ref.~\onlinecite{2009-Kambly}.

\medskip

%%%%%%%%%%%%%%%%
\begin{figure}[t]
\includegraphics[width=5cm]{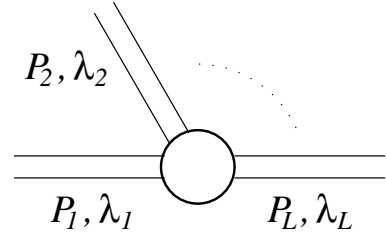}
\caption{A schematic figure of the multiterminal contact. 
To each of the $L$ leads, there corresponds a single-particle
projector $P_i$ and a counting variable $\lambda_i$.}
\label{fig:setup}
\end{figure}
%%%%%%%%%%%%%%%%

{\it Determinant formula. }
We first introduce notation and review the Levitov--Lesovik determinant
formula\cite{1993-LevitovLesovik,1993-IvanovLevitov,%
1996-LevitovLeeLesovik,1997-IvanovLeeLevitov,2002-Klich} 
for charge transfer of non-interacting fermions in 
application to a multi-lead setup. The notation and argument is fully
parallel to that in Ref.~\onlinecite{2009-AbanovIvanov} where the
two-lead case was considered.

We consider a contact with $L$ leads, connected by an arbitrary 
time-dependent scatterer (see Fig.~\ref{fig:setup}). To each lead 
(numbered $i=1,\ldots,L$) we associate a 
``counting field''\cite{1993-LevitovLesovik} $\lambda_i$
and a projector operator $P_i$ acting in the single-particle Hilbert
space. The leads are defined in such a way that
\begin{equation}
\sum_{i=1}^{L} P_i = \mathbf{1}\, .
\label{projectors}
\end{equation}
Then the probabilities of the multi-lead charge transfers can be
determined from the generating function
\begin{equation}
\chi(\lambda_1,\ldots,\lambda_L)=
\Tr\left(\hat{\rho}_{0} \hat{U}^{\dagger}
e^{i\mathbf{\lambda}\mathbf{\hat{P}}}\hat{U}e^{-i\mathbf{\lambda} 
\mathbf{\hat{P}}}\right)\Big/\,  
\Tr\hat{\rho}_{0} \, .
\label{FCS-1}
\end{equation}
Here the trace is taken in the multi-particle Fock space, 
$\hat{\rho}_{0}$ is the initial density matrix, $\hat{U}$ is the
multi-particle evolution operator. We also use the shorthand notation
$\mathbf{\lambda}\mathbf{\hat{P}} = \sum_i \lambda_i \hat{P}_i$, where
$\hat{P}_i$ is the multi-particle operator 
(a fermionic bilinear \cite{2009-AbanovIvanov}) constructed from the projector
$P_i$ (it counts the particles in the lead $i$). As in the two-lead
problem,\cite{2009-AbanovIvanov} under the assumption that 
$\hat{\rho}_{0}$ commutes with $\hat{P}_i$ (the absence of entanglement
in the initial state), the Fourier components
of the generating function (\ref{FCS-1}) give the charge-transfer
probabilities $P_{q_1,\ldots,q_L}$,
\begin{equation}
\chi(\lambda_1,\ldots,\lambda_L) = 
\sum_{q_1,\ldots,q_L=-\infty}^{\infty}P_{q_1,\ldots,q_L}\, 
\exp\left(i \sum_{i=1}^L \lambda_i q_i\right)\, .
 \label{FCS-2}
\end{equation}
Those probabilities are only non-zero for charge-conserving transfers
with $\sum_i q_i =0$. This charge conservation corresponds to the
symmetry of the generating function with respect to a simultaneous
shift of all variables,
\begin{equation}
\chi(\lambda_1,\ldots,\lambda_L) = 
\chi(\lambda_1 + \delta\lambda,\ldots,\lambda_L + \delta\lambda)\, .
\end{equation}
As in the two-lead case, we define the complex variables
\begin{equation}
u_i = e^{i\lambda_i}\, , \qquad i=1,\ldots,L \, ,
\end{equation}
and consider the generating function as a function of $u_i$.

As in Ref.~\onlinecite{2009-AbanovIvanov}, we assume, in addition to the
absence of entanglement of the initial state, that both $\hat{\rho}_{0}$
and $\hat{U}$ are exponentials of fermionic bilinears (which reflects our
assumption of non-interacting fermions). Under those assumptions, we
repeat the calculation of Ref.~\onlinecite{2009-AbanovIvanov} and
arrive at the resulting determinant formula
\begin{equation}
\chi(\lambda_1,\dots,\lambda_L)= 
\det\Big[1+n_{F}(U^{\dagger}e^{i\mathbf{\lambda}\mathbf{P}}U
e^{-i\mathbf{\lambda}\mathbf{P}}-1)\Big]\, ,
 \label{spFCS}
\end{equation}
which involves only operators in the single-particle Hilbert space
with the occupation-number operator
\begin{equation}
n_{F}=\frac{\rho_{0}}{\rho_{0}+1}\, .
\label{nF}
\end{equation}

\medskip

\begin{figure}[t]
\includegraphics[width=7cm]{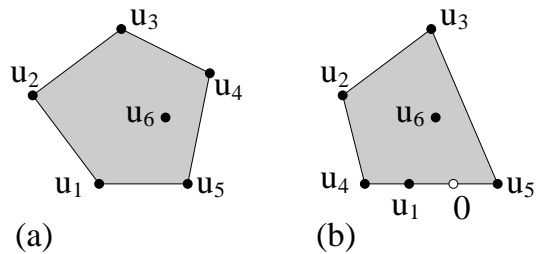}
\caption{{\bf (a):} Illustration of the definition of the convex
envelope (convex hull). The shaded region shows the convex envelope of the
points $u_1, \dots, u_6$ in the complex plane. If the points $u_1, \dots, u_6$
correspond to a root of the generating function, then Condition 1 of the
constraint claims that zero must belong to the shaded region.
{\bf (b):} Illustration of Condition 2 of the constraint. In this figure
(with the points $u_1, \dots, u_6$ corresponding to a root of the generating
function), the points $u_2$, $u_3$, and $u_6$ can be changed arbitrarily, and
the new set of points will still give a root of the generating
function.}
\label{fig:hull}
\end{figure}

{\it Convexity condition. }
Similarly to the trick employed in Ref.~\onlinecite{2009-AbanovIvanov},
we can rewrite the determinant formula by defining the
hermitian ``effective-transparency operators''
\begin{equation}
\tilde{X}_{(i)} = (1-n_{F})P_i +n_{F}^{1/2} U^{\dagger}P_i U n_{F}^{1/2} \, .
 \label{X}
\end{equation}
After simple algebra [using the completeness relation (\ref{projectors})], 
one can re-express the generating function (\ref{spFCS}) as
\begin{equation}
\chi(u_1,\dots,u_L) = \det\left[ e^{-i\mathbf{\lambda}\mathbf{P}}\,
\sum_{i=1}^L u_i \tilde{X}_{(i)} \right]\, .
\label{chi-X}
\end{equation}
The eigenvalues of the operators $\tilde{X}_{(i)}$ are bounded 
between 0 and 1, which
allows us to prove a certain constraint on the zeroes (roots)
of the generating
function (\ref{chi-X}). An elegant form of this constraint can be formulated
in terms of the {\em convex envelope} (convex hull) $H_c(X)$
of a given set of complex numbers $X$: a minimal convex set containing $X$
(see Fig.~\ref{fig:hull}a).
The constraint may now be cast in the form of two conditions that
need to be satisfied:
\begin{enumerate}
\item
For any root of the characteristic function $\chi(u_1,\dots,u_L)=0$,
the convex envelope $H_c(\{u_1,\dots,u_L\})$ contains zero.
\item
If $\chi(u_1,\dots,u_L)=0$ and if zero belongs to the {\em boundary} \
of $H_c(\{u_1,\dots,u_L\})$, then those of the points $\{u_1,\dots,u_L\}$
that do not lie on the straight segment of the boundary of
 $H_c(\{u_1,\dots,u_L\})$ containing zero, can be arbitrarily changed
while still satisfying the equation $\chi(u_1,\dots,u_L)=0$ 
(Fig.~\ref{fig:hull}b).
\end{enumerate}

%%%%%%%%%%%%%%%%
%%%%%%%%%%%%%%%%

The proof of Condition 1 is easy: if $|\Psi\rangle$ is a zero mode
of the operator in the determinant (\ref{chi-X}), then 
\begin{equation}
\sum_{i=1}^{L} u_i \langle \Psi | \tilde{X}_{(i)} | \Psi \rangle =0\, .
\label{linear-combination}
\end{equation}
Since all the coefficients $\langle \Psi | \tilde{X}_{(i)} | \Psi \rangle$
are non-negative real numbers (whose sum equals one), zero belongs to the
convex envelope of $u_1,\dots,u_L$.

To prove Condition 2, consider again a root $(u_1,\dots,u_L)$ of the
generating function and the corresponding zero mode $|\Psi\rangle$.
If zero lies at the boundary of the convex envelope  $H_c(\{u_1,\dots,u_L\})$,
then the linear combination (\ref{linear-combination}) contains
nonvanishing coefficients  $\langle \Psi | \tilde{X}_{(i)} | \Psi \rangle$
{\em only} for variables $u_i$ which belong to the same straight segment
of the boundary containing zero. All the other coefficients necessarily
vanish, which, by virtue of the non-negativity of $\tilde{X}_{(i)}$, implies
$\tilde{X}_{(i)} | \Psi \rangle = 0$. Therefore all those variables $u_i$ may
be changed arbitrarily while $| \Psi \rangle$ will remain a zero mode.
This completes the proof of Condition 2.

%%%%%%%%%%%%%%%%
\begin{figure}[t]
\includegraphics[width=5cm]{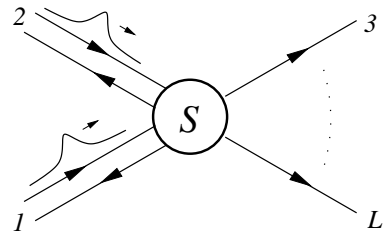}
\caption{A counterexample proving non-factorizability of the
full counting statistics for multi-terminal contacts. Two
fermions are sent to a time- and energy-independent $L$-lead
scatterer (with $L\ge 3$) along the leads 1 and 2 in the shape of
exactly identical wave packets, synchronized in time.}
\label{fig:counterexample}
\end{figure}
%%%%%%%%%%%%%%%%

We can make several comments on the obtained result. First, in the
particular case of two leads ($L=2$), this constraint is equivalent to
that found in Ref.~\onlinecite{2009-AbanovIvanov} (the variable $u$ in that
work corresponds to the ratio $u_1/u_2$ in our present notation). Second,
while our constraint is a necessary condition for realizability of a given
statistics in a non-interacting fermionic system, we could not determine 
if it is also a sufficient one. Moreover, we do not have any algorithm which
would determine if a given charge-transfer statistics is realizable (or
design a suitable quantum evolution if it is). Those interesting questions
are left for future studies. Third, our criterion is technically difficult
to check in its full formulation for all roots $(u_1,\dots,u_L)$. However,
for practical applications, one may test the constraint on suitably chosen
families of roots (e.g., one-parametric families\cite{test}), either 
analytically or numerically.

\medskip

{\it Non-factorizability. } 
In the two-terminal case, the ``convexity condition'' derived above implies
a factorizability of the charge transfer statistics: the probabilities of
a given charge transfer are the same as in a superposition of some 
single-electron transfer processes (whose transfer probabilities
depend in a non-trivial way on the evolution of the quantum system).
One can see that it is not the case in the multi-terminal ($L>2$) case.

This can be most easily demonstrated with a counter-example involving only
a finite number of electrons (in the wave packet formalism of 
Ref.~\onlinecite{2008-Hassler}, to which our result is also applicable).
Consider two fermions sent into a stationary multi-terminal contact 
along two terminals (labeled 1 and 2) with exactly the same 
shape of wave packets  (Fig.~\ref{fig:counterexample}). 
Then, due to the Fermi statistics of particles, the probabilities
to have both fermions scattered to the same lead vanish. The
resulting generating function will therefore have the form
\begin{equation}
\chi(u_1,\ldots,u_L) = \frac{1}{u_1 u_2}\sum_{i<j} \alpha_{ij} u_i u_j\, ,
\label{antisymmetric}
\end{equation}
where $\alpha_{ij} = | s_{1i} s_{2j} - s_{2i} s_{1j} |^2$ are the probabilities
of various two-particle transfer events constructed out of the 
single-particle scattering amplitudes $s_{ij}$ (which are assumed to
be time and energy independent). On the other hand, the factorizability
of the charge transfer would imply
\begin{equation}
\chi(u_1,\ldots,u_L) = \frac{1}{u_1 u_2} 
\left( \sum_i p_i u_i \right)
\left( \sum_i p'_i u_i \right)
\label{factorizable}
\end{equation}
for some probabilities $p_i$ and $p'_i$. One can verify that 
if one considers a statistics (\ref{antisymmetric}) with all $\alpha_{ij}$
nonzero (which is possible), then such a statistics is not factorizable
in the form (\ref{factorizable}).

{\it Conclusion. }
To summarize, we have considered the problem of possible full counting
statistics for non-interacting fermions in coherent multi-terminal systems. 
We have obtained a necessary condition for a full counting statistics
to be realizable. Like in the two-terminal case,\cite{2009-AbanovIvanov}
this condition may be used to prove impossibility of certain sets of
charge-transfer probabilities (one can easily construct examples of
such impossible statistics). 

At the same time, the problem of designing an actual ``quantum pump''
for a given charge-transfer statistics (or even merely proving 
its {\it possibility}) appears much more difficult in the multi-terminal
case than in the two-terminal one. While in the two-terminal case,
the full counting statistics of non-interacting fermions is conveniently
parameterized by the spectral density of 
``effective transparencies'',\cite{2009-AbanovIvanov} we are not aware of
a similar parameterization in the multi-terminal case. 
In the formulation with a finite number of particles\cite{2008-Hassler},
even the question of the dimensionality of the space of all possible
full counting statistics remains open. All those interesting questions
deserve further study, in particular in the context of using
quantum contacts for generating entangled states.\cite{2008-KlichLevitov}

%%%%%%%%%%%%%%%%%%%%%%%%%%%%%%%%%%%%%%%%

%%%%%%%%%%%%%%%%%%%%%%%%%%%%%%%%


\begin{thebibliography}{99}
\bibitem{test}
A good practical test of the constraint is obtained by restricting $u_i$ to
a one-parameter family $u_i = a_i z +b_i$ for some fixed sets of real
numbers $a_i$ and $b_i$ with the condition that all $a_i \ge 0$ 
(nothing is assumed about $b_i$). Then the equation $\chi(z)=0$ must
either have only real roots $z$ or be identically satisfied for all $z$.
This form of the constraint is convenient for numerical tests
by using a large number of randomly chosen  vectors $a_i$ and $b_i$.
\end{thebibliography}
\end{document}